\documentclass[a4paper,11pt]{article}

\usepackage[margin=1.14in]{geometry} 

\usepackage{amsmath}
\usepackage{amsthm}
\usepackage{amssymb}
\makeatletter
\makeatletter \renewcommand{\@dotsep}{10000} \makeatother
\usepackage[utf8]{inputenc}
\usepackage{lmodern}
\usepackage{hyperref}
\hypersetup{
	unicode,
	colorlinks,
	breaklinks,
	urlcolor=cyan, 
    linkcolor=black, 
	pdfauthor={Author One, Author Two, Author Three},
	pdftitle={A simple article template},
	pdfsubject={A simple article template},
	pdfkeywords={article, template, simple},
	pdfproducer={LaTeX},
	pdfcreator={pdflatex}
}
\usepackage[T1]{fontenc} % if needed
\usepackage{amsmath}

\newcommand{\be}{\begin{eqnarray}}
\newcommand{\ee}{\end{eqnarray}}
\def\be{\begin{equation}}
\def\ee{\end{equation}}
\def\bea{\begin{eqnarray}}
\def\eea{\end{eqnarray}}

\newcommand{\gsim}{\;\raisebox{-0.9ex}{$\textstyle\stackrel{\textstyle >}{\sim}$}\;}
\newcommand{\lsim}{\;\raisebox{-0.9ex}{$\textstyle\stackrel{\textstyle<}{\sim}$}\;}
\def\lsim{\raise0.3ex\hbox{$\;<$\kern-0.75em\raise-1.1ex\hbox{$\sim\;$}}}
\def\gsim{\raise0.3ex\hbox{$\;>$\kern-0.75em\raise-1.1ex\hbox{$\sim\;$}}}

\usepackage{graphics}

\usepackage{epsfig}
\usepackage{slashed}
\usepackage[utf8]{inputenc}
\newcommand\Tstrut{\rule{0pt}{2.6ex}}         % = `top' strut
\newcommand\Bstrut{\rule[-0.9ex]{0pt}{0pt}}   % = `bottom' strut
\usepackage{multirow}
\usepackage{pstricks}
\usepackage{dcolumn}

\usepackage[sort&compress,numbers,square]{natbib}
\theoremstyle{plain}

\theoremstyle{definition}

\usepackage{graphicx, color}

\usepackage[linesnumbered,ruled,vlined,commentsnumbered]{algorithm2e} % use algorithm2e for typesetting algorithms
\usepackage{mathrsfs} % for \mathscr command

\title{Search for Mono-Higgs Signals in $b\bar b$ Final States\\[0.25cm]
Using Deep Neural Networks}
\author{\Large{A. Hammad$^a$\,, S. Khalil$^b$ and S. Moretti$^{c,}$$^d$}}
\date{
$^a$ Institute of Convergence Fundamental Studies, Seoul National University of
Science and Technology, 232 Gongneung-ro, Nowon-gu, Seoul, 01811, Korea. \\ 
$^b$ Center for Fundamental Physics, Zewail City of Science and Technology, 6 October City, Giza 12588, Egypt.\\ 
$^c$ School of Physics and Astronomy, University of Southampton,
Highfield, Southampton SO17 1BJ, UK.\\
$^d$ Department of Physics $\&$ Astronomy, Uppsala University, Box 516, SE-751 20 Uppsala, Sweden.
}

\begin{document}
	\maketitle
	\vspace{4mm}
	\begin{abstract}
 \normalsize{We study mono-Higgs signatures emerging in an illustrative new physics scenario involving  Standard Model Higgs boson decays to bottom quark pairs using Hybrid Deep Neural Networks. We use a Multi-Layer Perceptron to analyze the kinematic observables and optimize the signal-to-background discrimination. The global color flow structure of hard jets emerging from the decay of heavy particles with different color charges is crucial to single out the mono-Higgs signature. Upon embedding the different color flow structures for signal and backgrounds into constructed images, we use a Convolution Neural Network to analyze the latter. Specifically, the approach  takes initially a mono-type data as input, frittering away invaluable multi-source and multi-scale information.  We then discuss a general architecture of Hybrid Deep Neural Networks   that supports instead mixed input data. In comparison with single input Deep Neural Networks, like MultiLayers Perceptron or Convolution Neural Network, the  Hybrid Deep Neural Networks provide higher capacity in  feature extraction  and thus in signal vs background classification performance. We provide reference results for the case of the High-Luminosity Large Hadron Collider. }  
\end{abstract}
\newpage
\noindent\rule{\textwidth}{1pt}
\tableofcontents
\noindent\rule{\textwidth}{0.2pt}
\maketitle \flushbottom
\vspace{4mm}
%%%%%%%%%%%%%%%%%%%%%%%%%%%%%%%%%%%%%%%%%%%%%%%%%%%%%%%%%%%%%%%%%%%%%%%%%%
\section{Introduction}
\label{sec:intro}
%%%%%%%%%%%%%%%%
Mono-$X$ is a signal of a single particle produced alongside Missing $E_T$ (MET), where $E_T$ represents transverse energy missing (transverse) energy, emerging at hadron colliders, where $X$ stands generally for a (light) jet, $W^\pm$, $Z$ or  photon. In recent years, these processes have received a lot of attention as a way to investigate new physics Beyond the Standard Model (BSM). In particular, mono-$X$ signatures of the above type would provide an indirect hint for Dark Datter (DM), which would indeed manifest itself as MET. A common question is which mono-$X$ channel works best. The mono-jet one has a large cross section relatively to the other modes, but the probe (a gluon or a quark) does not typically couple to DM or the mediator. Also, such a channel can suffer from large Quantum Chromo-Dynamics (QCD) backgrounds. Conversely, the other probes mentioned above could {\sl also} couple to DM and/or the mediator in presence of a smaller background but they are very suppressed as they are Electro-Weak (EW) processes in nature. Altogether, to date, most sensitivity is acquired via mono-jet processes. 

In the above list, we have purposely left aside another possible mono-$X$ probe, where $X$ is a Higgs boson. This could be the SM-like one discovered at the Large Hadron Collider (LHC) or else a companion state emerging in many BSM scenarios. We are interested in this paper in establishing a mono-Higgs signal, wherein such a state decays into a $b\bar b$ pair. The latter is the most frequent decay channel for a light (neutral) Higgs boson, including the SM-like one discovered in 2012 (hereafter denoted by $h$), 
so it is of phenomenological relevance. Also this decay, however, requires one to deal with an overwhelming QCD background, so that attempts to extract it in mono-Higgs events have been not very fruitful to date, including in terms of eventually establishing sensitivity to DM and/or the mediator properties, in comparison to what can be established through the other aforementioned mono-$X$ channels. This is clearly a pity,
as mono-Higgs has a distinctive advantage with respect to all other mono-$X$ channels, i.e.,  in proton-proton collisions, for the cases $X={\rm jet}, W^\pm, Z$ and $\gamma$, the probe can always be emitted directly from a light quark as Initial State Radiation (ISR) through the  usual SM gauge interactions even when the $X$ object interacts with the remainder of the process. In contrast, ISR induced by Higgs-strahlung (i.e., mono-Higgs) is highly suppressed due to the small
coupling of the $h$ state to light quarks. Hence, unlike other mono-$X$ signatures, a mono-Higgs one would probe exclusively the properties of DM and/or the mediator.

Therefore, here, we take a step in the direction of improving the scope of such a signature, i.e., the mono-Higgs one with $h\to b\bar b$, by looking at signal events which have a peculiar kinematics with respect to the background. The opportunity is offered by the $B-L$ Supersymmetric Standard Model (BLSSM) of Refs.~\cite{{Khalil:2007dr}}, wherein heavy BSM Higgs states (with masses up to around TeV) can mediate final states involving invisible DM and a Higgs boson \cite{Abdallah:2016vcn}. In this case, the final state contains a large amount of MET, allowing the BLSSM signature to be accessed above the SM background. However, in \cite{Abdallah:2016vcn}, only the $\gamma\gamma$ and $ZZ^*\to 4l$ ($l=e,\mu$) decays of the SM-like Higgs were established, as the $b\bar b$ one was found to be overwhelmed by QCD radiation yielding $b\bar b$ pairs, upon adopting a signal-to-background analysis solely based on kinematic cuts. 

It is the purpose of this paper to revisit such an analysis, with the intent of
extracting the mono-Higgs signal $h\to b\bar b$ in presence of Machine Learning (ML) techniques.  The reason for doing this is twofold. On the one hand, since the appearance of \cite{Abdallah:2016vcn}, the $h\to b\bar b$ decay has now been fully established (and found to be compliant with SM predictions \cite{ATLAS:2018kot,CMS:2018nsn}), so that one can enforce the $h$ mass reconstruction from the $b\bar b$ system. On the other hand, actual experimental searches based on standard approaches  have primarily exploited alternative $h$ decays, for the aforementioned reasons: e.g., the recent ATLAS \cite{ATLAS:2021shl,ATLAS:2021jbf} and CMS \cite{CMS:2019ykj,CMS:2018nlv} analyses covered the  mono-Higgs channels with $h$ decays to $ZZ+$MET, $W^+W^- +$MET, $\gamma\gamma+$MET and $\tau^+\tau^- +$MET, all of which have 
better sensitivity than the $b\bar b$+MET channel (and reporting no significant excess over the expected SM
background). 

Here, we surpass the state-of-the-art in mono-Higgs searches with $h\to b\bar b$ as
we exploit three independent Deep Neural Networks (DNNs). Firstly, we invoke Multi-Layers Perceptron (MLP) systems, which analyze the constructed kinematic distributions of the final state particles. Secondly, we use Convolution Neural Networks (CNNs), which analyze the jet images that can be constructed by embedding the $p_T$ distribution of the final state jets into visual representations, wherein the different color flow structure of the signal and background processes is accounted for. Thirdly, we deploy Hybrid Deep Neural Network (HDNN), which are a two-stream input framework that can analyze the kinematic distributions and constructed jet images at the same time, thereby being able to enhance the signal-to-background classification performance better than the previous approaches.

This paper is organized as follow.   In section \ref{sec:2} we discuss the  mono-Higgs signals via $h\to b\bar b$ arising in the BLSSM, after a very brief discussion of the latter. In section \ref{sec:3} we discuss our analysis strategy. Section \ref{sc:4} shows the described network architectures and their data pre-processing stages. The results of our analysis,
exemplified for the High-Luminosity LHC (HL-LHC) \cite{Gianotti:2002xx}, are given in section \ref{sc:5}. Finally, we summarize and  conclude in section \ref{sc:6}.

%%%%%%%%%%%%%%%%%%%%%%%%%%%%%%%%%%%%%%%%%%%%%%%%%%%%%%%%%%%%%%%% Sec. II %%%%%%%%%%%%%%%%%%%
\section{Mono-Higgs in the BLSSM} \label{sec:2}
In addition to the (s)particles of the Minimal Supersymmetric Standard Model (MSSM), the BLSSM  includes three chiral
right-handed superfields ($\hat{N}_i$), a vector superfield associated
to $U(1)_{B-L}$ ($\hat{Z}'$) and two chiral SM singlet Higgs
superfields ($\hat{\eta}_1$, $\hat{\eta}_2$), as discussed in details in , with the following superpotential:
\bea
\hat{W} = Y_u \hat{Q}\hat{H}_2\hat{U}^c + Y_d\hat{Q}\hat{H}_1\hat{D}^c + Y_e\hat{L}\hat{H}_1\hat{E}^c +Y_{\nu}\hat{L}\hat{H}_2\hat{N}^c + Y_N\hat{N}^c\hat{\eta}_1\hat{N}^c+\mu \hat{H}_1\hat{H}_2 + \mu'\hat{\eta}_1\hat{\eta}_2.\nonumber
\eea
The quantum numbers for the particles content in the BLSSM are given in Tab. \ref{particle-content-BLSSM}
\begin{table}[h!]
\begin{center}
\begin{tabular}{|c||c|c|c|c|c|c|c|c|c|c|}\hline\hline
& $\hat{Q}_i$ & $\hat{U}_i^c$ & $\hat{D}_i^c$ & $\hat{\ell}_i$ & $\hat{E}_i^c$ & $\hat{N}_i^c$ & $\hat{H}_1$ & $\hat{H}_2$ & $\hat{\eta}_1$ & $\hat{\eta}_2$  \Tstrut\Bstrut \\
\hline\hline
$SU(3)_C$ & $3$ & $\bar{3}$ & $\bar{3}$ & $1$ & $1$ & $1$ & $1$ & $1$ & $1$ & $1$\\
\hline
$SU(2)_L$ & $2$ & $1$ & $1$ & $2$ & $1$ & $1$ & $2$ & $2$ & $1$ & $1$\\
\hline
$U(1)_Y$ & $1/6$ & $-2/3$ & $1/3$ & $-1/2$ & $1$ & $0$ & $-1/2$ & $1/2$ & $0$ & $0$\\
\hline
$U(1)_{B-L}$ & $1/6$ & $1/6$ & $1/6$ & $-1/2$ & $1/2$ & $1/2$ & $0$ & $0$ & $-1$ & $1$\\
\hline\hline
\end{tabular}
\caption{Particle content of the BLSSM}
\label{particle-content-BLSSM}
\end{center}
\end{table}

After $ B-L $ and EW symmetry breaking, the following mass for the BLSSM-like CP-odd Higgs $A'$ is obtained: 
\be
m_{A'}^{{2}} =\frac{2 B {\mu'}}{\sin2 \beta'} \sim {\cal O}(1~{\rm TeV}),
\ee 
where $\tan\beta'=v'_1/v'_2$. The BLSSM CP-even neutral Higgs fields have the following masses at tree level:
\bea
{m}^2_{h',H'} &=& \frac{1}{2} \Big[ ( m^2_{A'} + M_{{Z'}}^2 )\mp \sqrt{ ( m^2_{A'} + M_{{Z'}}^2 )^2 - 4 m^2_{A'} M_{{Z'}}^2 \cos^2 2\beta' }\;\Big].
\eea
If $\cos^2{{2}\beta'} \ll 1$, one finds that the lightest $B-L$ neutral Higgs mass is given by %
\be%
{m}_{h'}\; {\simeq}\; \left(\frac{m^2_{A'} M_{{Z'}}^2 \cos^2 2\beta'}{{m^2_{A'}+M_{{Z'}}^2}}\right)^{\frac{1}{2}} \simeq {\cal O}(100~ {\rm GeV}).%
\ee%

Another important sector of the BLSSM spectrum is the neutralino one. The  neutral
gaugino-higgsino mass matrix can be written as \cite{Khalil:2007dr}: %
\bea
{\cal M}_7({\tilde B},~{\tilde W}^3,~{\tilde
H}^0_1,~{\tilde H}^0_2,~{\tilde B'},~{\tilde \eta_1},~{\tilde
\eta_2}) \equiv \left(\begin{array}{cc}
{\cal M}_4 & {\cal O}\\
 {\cal O}^T &  {\cal M}_3\\
\end{array}\right),\nonumber
\eea%
where  ${\cal M}_4$ is the MSSM-type neutralino mass matrix \cite{Haber:1984rc,Gunion:1984yn,ElKheishen:1992yv,Guchait:1991ia} and
${\cal M}_{3}$ is $3\times 3$ additional neutralino mass matrix,
which is given by%
\be%
{\cal M}_3 = \left(\begin{array}{ccc}
M_{B'} & -g_{_{B-L}}v'_1  & g_{_{B-L}}v'_2 \\
-g_{_{B-L}}v'_1 & 0 & -\mu'  \\
g_{_{B-L}}v'_2 & -\mu' & 0\\
\end{array}\right).
\label{mass-matrix.1} \ee
In addition, the off-diagonal matrix ${\cal O}$ is given by
\be%
{\cal O} = \left(\begin{array}{ccc}
\frac{1}{2}M_{BB'} &~~~0~~~& 0 \\
0 & 0 & 0  \\
-\frac{1}{2}\tilde{g}v_1 &~~~0~~~& 0\\
\frac{1}{2}\tilde{g}v_2&~~~0~~~&0\\
\end{array}\right).
\label{mass-matrix.1} \ee
Note that the off-diagonal matrix elements herein vanish identically if $\tilde{g}=0$. In this case, one diagonalises the real matrix ${\cal M}_{7}$ with
a symmetric mixing matrix $V$ such that
%%%
\be V{\cal
M}_7V^{T}={\rm diag}(m_{\tilde\chi^0_k}),~~k=1,\dots, 7.\label{general} \ee In
these conditions, the Lightest Supersymmetric Particle (LSP), i.e., the DM candidate of the BLSSM, has the following decomposition 
\bea 
\tilde\chi^0_1&=&V_{11}{\tilde B}+V_{12}{\tilde
W}^3+V_{13}{\tilde H}^0_1+V_{14}{\tilde
H}^0_2 +V_{15}{\tilde B'}+V_{16}{\tilde \eta_1}+V_{17}{\tilde
\eta_2}. 
\eea 
The LSP is called pure
$\tilde B'$ if $V_{15}\sim1$ and $V_{1i}\sim0$ for $i\neq5$ and pure $\tilde\eta_{1(2)}$ if $V_{16(7)}\sim1$
and all the other coefficients are close to zero. 

\begin{figure}[h!]
\begin{center}
\includegraphics[width=4cm,height=4.cm]{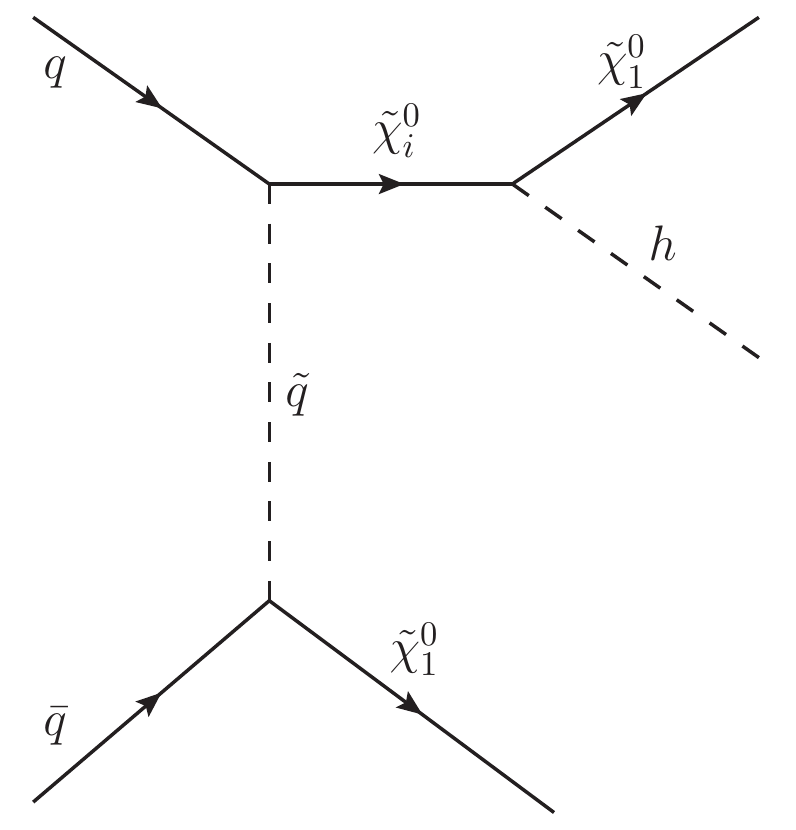}\includegraphics[width=4.7cm,height=3.cm]{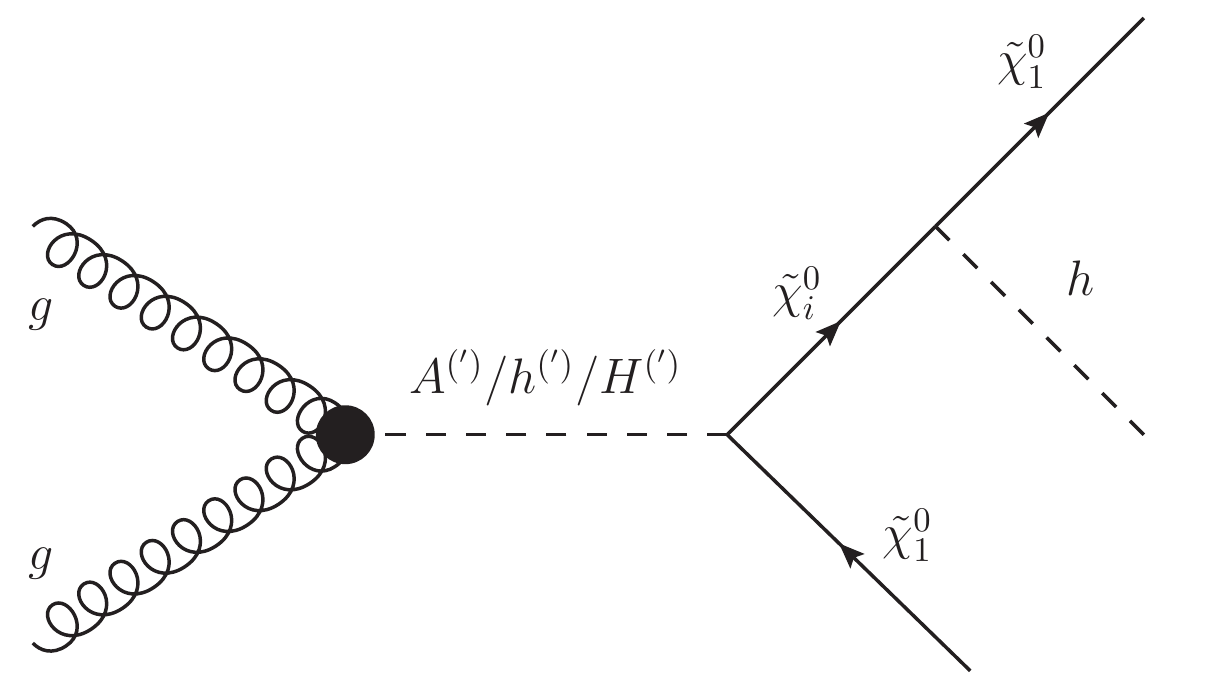}
\includegraphics[width=4.7cm,height=4.cm]{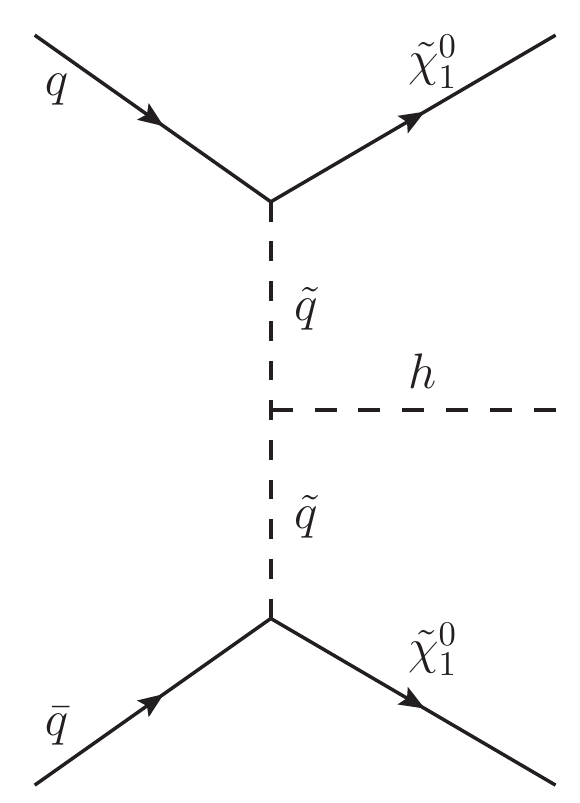}
\caption{Mono-Higgs Feynman diagrams $q\bar{q}\to\tilde{\chi}_1^0\tilde{\chi}_i^0\to\tilde{\chi}_1^0\tilde{\chi}_1^0 h$ with $\tilde{q}$ exchange
(left diagram),
$gg \to A^{(')}/h^{(')}/H^{(')}\to\tilde{\chi}_1^0\tilde{\chi}_i^0\to\tilde{\chi}_1^0\tilde{\chi}_1^0 h$ (middle diagram) and  $q\bar{q}\to\tilde{\chi}_1^0\tilde{\chi}_i^0\to\tilde{\chi}_1^0\tilde{\chi}_1^0 h$ with $\tilde{q}\tilde{q}$ exchange.
(right diagram)}
\label{fig:Feynman}
\end{center}
\end{figure}

Now, we can investigate the Mono-Higgs signal of concern in such a BLSSM model. This signal  can be produced as initial, intermediate or final state $h$-radiation associated with DM pair production (i.e., $\tilde\chi^0_1\tilde\chi^0_1$) \cite{Abdallah:2016vcn}.  Representative Feynman diagrams for the considered signal processes are shown in figure \ref{fig:Feynman}. (Notice that we ignore herein $Z'$ mediated topologies, 
considered in \cite{Abdallah:2016vcn}, as they are presently negligible, given the latest limits on $Z'$ massess and couplings in the BLSSM \cite{CMS:2021ctt}.)
The considered  mono-Higgs process benefits then from three sub-processes, two involve $t,u$ channel exchange of squarks, with the $h$ emerging from either a $\tilde q$ or a  $\tilde{\chi}_i^0$, and one includes $s$-channel propagation of all neutral Higgs boson states of the BLSSM, with the $h$ emerging from a $\tilde{\chi}_i^0$. 

In order to be consistent with the current experimental bounds we adopt a Benchmark Point (BP) from the scan performed in \cite{Abdallah:2016vcn} with the following features (as mentioned, we take $m_h=125$ GeV): 
\begin{equation*}
m_{{\tilde{\chi}^0}_1} = 48.4 ~\text{GeV} \,, \hspace{4mm} m_{h^\prime} = 263.8 ~\text{GeV}\,, \hspace{4mm}  m_{{\tilde{\chi}^\pm}_1} =712 ~\text{GeV}\,, \hspace{4mm} m_{A^\prime}\sim m_{H^\prime} = 800 ~\text{GeV} \,,
\end{equation*}
\begin{equation*}
m_{{\tilde{\chi}^0}_{2,3}} = \mathcal{O}(100~\text{GeV})\,, \hspace{4mm} m_{{\tilde{\chi}^0}_{4,5,6,7}} = \mathcal{O(\text{TeV})}\,, \hspace{4mm} m_{\tilde{q}}= \mathcal{O(\text{TeV})}\,, \hspace{4mm} m_{A}\sim m_H =\mathcal{O(\text{TeV})}\,. 
\end{equation*}
Thus, for our BP, the DM is significantly lighter than heavier neutralinos and  the neutral Higgs states, which will have implications for the event kinematics of the signal.
We have checked that the selected BP  is consistent with current  LHC Run-2 bounds and allowed by all the Higgs searches and measurements at the $95\%$ Confidence Level (CL) as checked by HiggsBounds-v5.3.2 \cite{Bechtle:2008jh,Bechtle:2011sb,Bechtle:2013wla} and HiggsSignals-v2.2.3 \cite{Bechtle:2013xfa,Stal:2013hwa,Bechtle:2014ewa}, respectively. Moreover, the LSP  satisfies the latest LUX bounds on the Direct Detection (DD) search for DM therein and all other experimental limits \cite{Akerib:2015rjg}. However, the relic abundance depends on the details of the underlying cosmology (thermal or non-thermal abundance), so its constraints will not be considered here \cite{Aprile:2015uzo,Kolb,Giudice:2000ex,Moroi:1999zb}.  

The total cross section for the combined signal topologies including SM-like Higgs decays to bottom quark pairs at $\sqrt{s}=14$ TeV is  $9\times 10^{-3}$ pb. Assuming 1000 fb$^{-1}$ of luminosity at the HL-LHC, this represents a sizable signal event sample, thereby affording one with the possibility of efficiently 
exploiting ML techniques.  

%
%%%%%%%%%%%%%%%%%%%%%%%%
\section{Analysis}\label{sec:3}
%%%%%%%%%%%%%%%%%%%%%%%%%
In this section, we describe the strategy behind our numerical analysis as well as the constituent elements of it, i.e.,  the event generation and detector simulation procedure as well as the signal and background properties in terms of kinematics and colour dynamics of jets. 

\subsection{Strategy}
 
 Having clarified the theoretical setup, 
  we now showcase a phenomenological study of mono-Higgs signals in $b\bar{b}$+MET final states using ML methods based on three independent DNNs. 

To start, global event information is extracted from constructing all relevant  kinematic distributions for signal and background events. A MLP model is then  adopted to optimize the separation power between the two  by analyzing such  distributions.  The fact that some background processes  have similar kinematic structures as the signal  hinders  the classification efficiency of this network, though. Moreover, the huge backgrounds cross section  leads to smaller signal significance even after optimizing the cut on the output score. To improve the MLP performance, one could then apply initial cuts on some variables before  feeding the distributions to the MLP to enhance the signal and   suppress the backgrounds. Because of the correlations amongst the constructed kinematic variables, applying a cut on some variables will affect all others, though, which in turn may undermine the network classification performance. To improve the impact of the initial cuts, one has then to de-correlate such a dependence across the kinematic variables via the square-root of the covariance matrix as in \cite{Hocker:2007ht}. However, in our analysis, we opted not to apply any such cuts on the constructed kinematic variables.

Instead, we adopted a second approach following the fact that $SU(3)$ of color is conserved in the  interaction processes and provides  different color flow structures for different processes. This depends on the color nature of the  interacting particles, e.g., the radiation pattern within and around the bottom-quark pair from Higgs boson decays is expected to be different from the radiation pattern  of bottom quark pairs from $t\bar{t}$ production and decay or prompt QCD processes. To quantify the impact on the signal-to-background rate of different color structures, one can think of a LHC  detector as a giant camera and the streams of hadrons as an image. Jet images can then be constructed as two dimensional arrays in the $(\eta, \phi)$ plane and the pixels can be weighted by the sum of the total transverse momentum deposited in the corresponding calorimetric region of the detector \cite{Komiske:2016rsd,Fraser:2018ieu,Cogan:2014oua,Almeida:2015jua,deOliveira:2015xxd}. We then  adopt a CNN model to analyze the constructed jet images in order to maximize the signal-to-background classification efficiency.

\begin{figure}[ht!]
\includegraphics[scale=0.47]{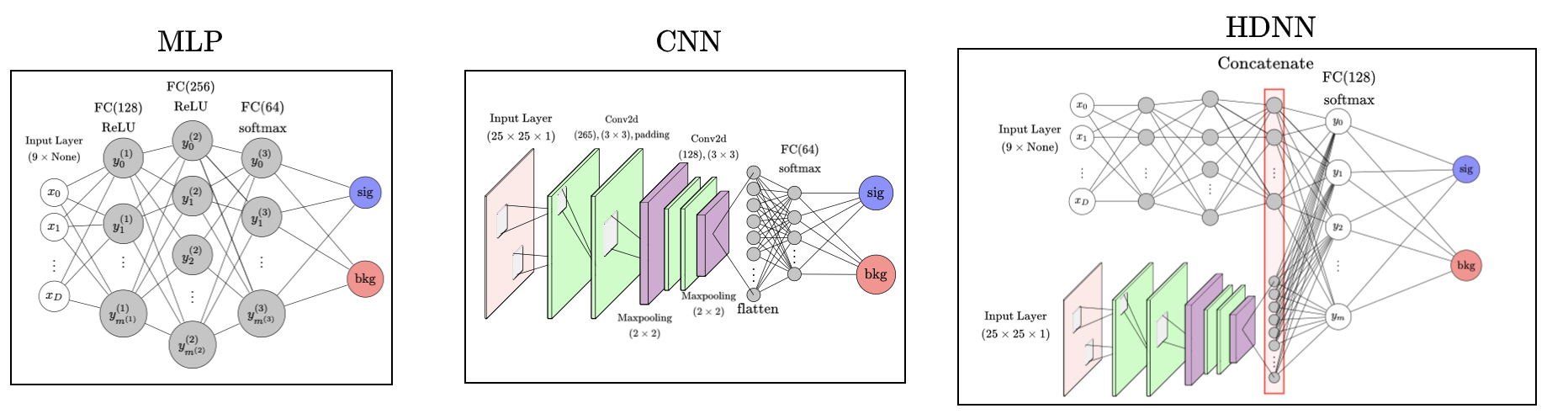}
\caption{A schematic architecture for the used DNNs: MLP (left), CNN (middle) and HDNN (right). Unlabeled layers of HDNN are similar to the ones in MLP and CNN.}
\label{fig:model}
\end{figure}

Finally, 
to incorporate the different data structures as inputs, whether kinematics or color flows, a two stream DNN is constructed \cite{Lin:2018cin,Kim:2019wns,Huang:2022rne}. The first stream, which processes the input jet images, consists of convolution, max-pooling and drop-out layers plus one flattened layer. The second stream, which processes the kinematic distributions, consists of fully connected  and drop-out layers. Both streams are then concatenated to one fully connected layer and one output layer with two neurons for predictions, hence, a Hybrid DNN (HDNN). 

Fig.  \ref{fig:model} shows the schematic architecture of the three DNNs: MLP (left), CNN (middle) and  HDNN (right).  
Feeding the HDNN model with different kinds of information enables it to extract the characteristic features of the signal and background events without requiring initial cuts or manual feature optimization. Moreover, combining the global color structure and the events kinematics boosts  the HDDN model performance   above and beyond that of the single stream networks, whether MLP or CNN, where individual streams are used. In the following, we discuss the structure of the three DNNs in detail and illustrate how to prepare the input data for the networks. Before doing so, though, we  describe how we have generated our Monte Carlo (MC) events.

%%%%%%%%%%%%%%%%%%%%%%%%%%%%%%%%%%%%%%%%%%%%%%%%%%%%%%%%
\subsection{Events generation and detector simulation}%%
%%%%%%%%%%%%%%%%%%%%%%%%%%%%%%%%%%%%%%%%%%%%%%%%%%%%%%%%
Both signal and background are simulated with MadGraph5 \cite{Madgraph5}, which is used to estimate multi-parton amplitudes and to generate events for  subsequent processing. All processes in Fig.~\ref{fig:Feynman} are computed at Leading Order (LO) 
except Higgs production from gluon-gluon fusion, which is calculated at Next-to-LO (NLO) in QCD using an effective coupling calculated by SPheno \cite{PorodSPheno,florianSPheno}. PYTHIA \cite{Sjostrand:2006za} is used for parton  showering, hadronization, heavy flavor decays and for adding the soft underlying event. The simulation of the response of, e.g., the ATLAS  detector was done with the DELPHES package \cite{deFavereau:2013fsa}. We slightly modified the standard ATLAS card therein to allow for the extraction of the track and energy deposit information for the final state hadrons. In fact,  reconstructed objects are simulated from the parametrized detector response and include tracks, calorimeter deposits as well as high level objects such as isolated electrons, jets, taus (both leptonic and hadronic) as well as MET.
%%%%%%%%%%%%%%%%%%%%%%%%%%%%%%%%%%%%%%%%%%%
\subsection{Signal and background kinematics}%
%%%%%%%%%%%%%%%%%%%%%%%%%%%%%%%%%%%%%%%%%%%

Considering the discussed BP and the signal topologies shown in Fig.  \ref{fig:Feynman}, the dominant SM background contributions arise from   the semi-leptonic decay of $t\bar t$ pairs, di-gauge boson production $pp \to hZ,  ZZ$ and, finally, vector boson production with bottom quark pairs stemming  from QCD radiation, $pp\to V b\bar{b}$ with $V= W\text{ or } Z$. We omit here  tri-gauge boson production which has a small cross section in comparison. Also, we neglect  the processes $pp \to b\bar{b}$ and $pp \to b\bar{b}+j$ with MET from the jet mis-reconstruction or from mesons decay, by assuming large missing energy cut, MET $\ge 250$ GeV \cite{Carpenter:2013xra,Bhowmik:2020spw}. 

 Based on the kinematics of signal and backgrounds, in order to generate the events
more effectively, we apply the following cuts at the simulation (i.e., MadGraph level): $p_T ({\rm jet})\ge 20$ GeV, $p_T (l)\ge 20$ GeV, $p_T (b)\ge 20$, MET $\ge 30$ GeV and  $|\eta (l, {\rm jet}, b)| \le 2.5$ (hereafter, $j=e,\mu$).   Moreover,  we require the events to have at least two $b$-jets with cone radius $R=0.4$\footnote{The analysis is insensitive to whether we use the $k_T$ \cite{Ellis:1993tq}, anti-$k_T$ \cite{Cacciari:2008gp, Catani:1993hr} or Cambridge-Aachen \cite{Dokshitzer:1997in,Wobisch:1998wt}  jet clustering algorithm.} using a flat tagging efficiency of $70\%$. (As for the mis-tagging efficiency of gluon- and light-quark-jets as $b$-ones, we adopt a flat rate of $10^{-3}$.) After generating the events we reconstruct nine kinematic distributions for both signal and backgrounds as detailed below. To collect the two $b$-jets from the Higgs boson decay we do not require any mass window cut rather we collect the $b$-jet pair that has the closest invariant mass to the SM-like Higgs boson mass. 

\begin{figure}[ht!]
\begin{center}
\includegraphics[scale=0.28]{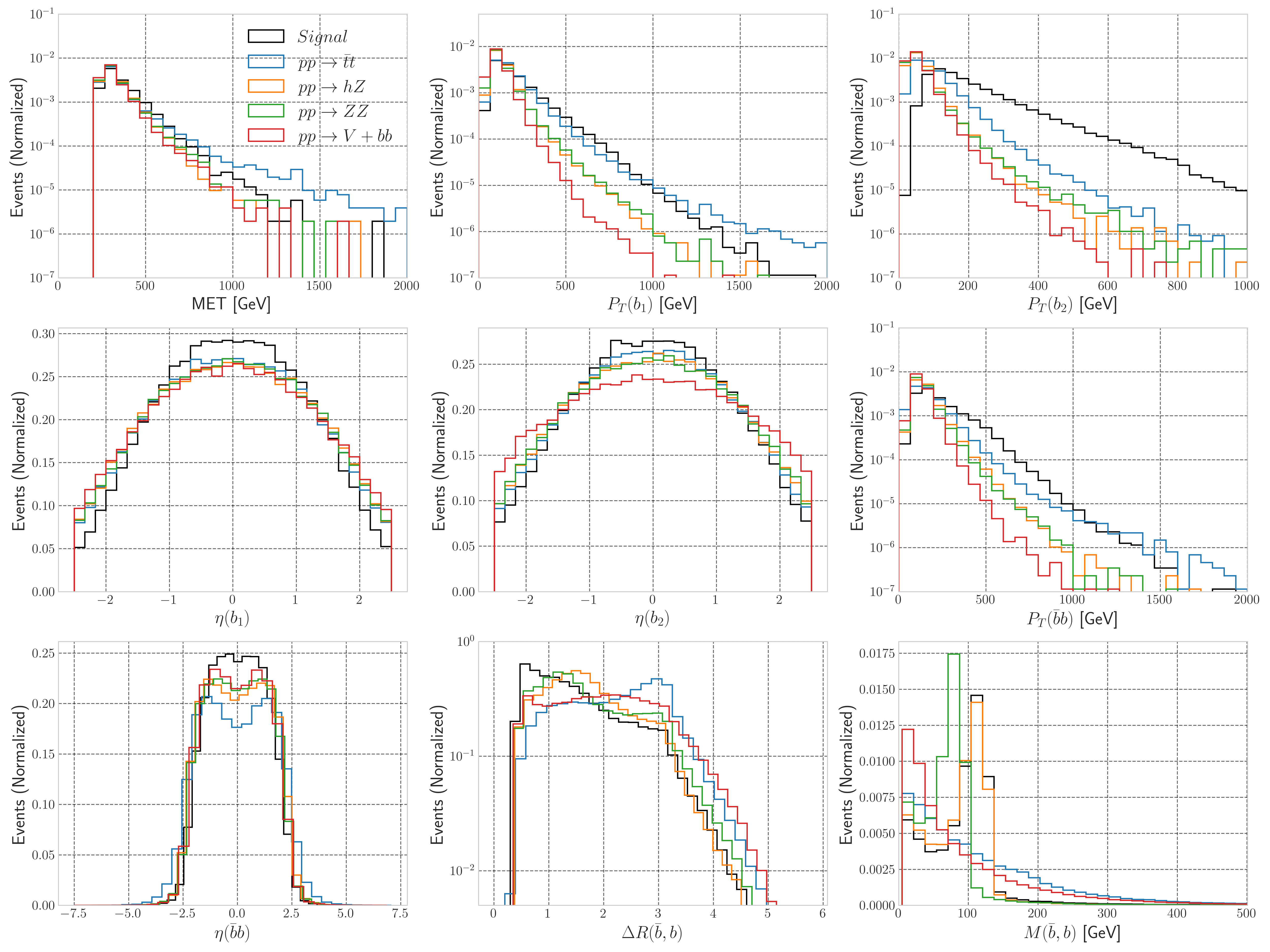}\\
\caption{The kinematic distributions discussed in the text for signal and background events superimposed and normalized to 1. The color codes apply for all distributions as follows: signal (black), $pp\to t\bar{t}$ (blue), $pp\to hZ$ (orange), $pp\to ZZ$ (green) and $pp\to Zb\bar  b$ (red).}
\label{fig:kinematics}
\end{center}
\end{figure}

Once  events have passed the selection criteria we exploit  the following kinematic variables, which we have plotted in Fig.~\ref{fig:kinematics}.
\begin{itemize}
\item  MET: Defined as ${\rm MET}
 =|-\sum_{v_i}\vec{p_T}(v_i)|$. Mono-Higgs signals with neutralino DM with mass about $50$ GeV allow for large MET that  can help to suppress the  $t\bar t$ and di-gauge boson backgrounds. 
\item \textbf{$p_T(b_1)$:} Transverse momentum of the leading $b$-jet originating from  SM-like Higgs boson decays. The rest mass of the (rather) heavy mediator Higgs boson (with mass of $\mathcal{O}(100$--$800)$ GeV, see Fig. \ref{fig:Feynman} (middle)), in comparison to the DM mass, boosts the final state $b$-jets showing a similar distribution to that of $t\bar{t}$ production and decay (wherein $b$-jets originate from top quark decays) while this kinematics is notably different   for the other backgrounds. 
\item \textbf{$p_T(b_2)$:} Transverse momentum of the second leading $b$-jet, which shows a similar behavior as $p_T(b_1)$ in both signal and background events.
\item \textbf{$\eta(b_1)$:} Pseudorapidity  of the leading $b$-jet, which shows a similar behavior in  both  signal and background events except for the process $pp\to Zb\bar  b$, showing a broader peak.
\item \textbf{$\eta(b_2)$:} Pseudorapidity  of the second leading $b$-jet, which shows a similar behavior as $\eta(b_1)$ for all processes.
\item \textbf{$p_T(b\bar{b})$:} Transverse momentum of the two  $b$-jets 
best reconstructing $m_h$ which exemplifies the Higgs boson boost in  signal events (again, signal and $t\bar{t}$ events have similar distributions which are in turn different from all other background processes). 
\item \textbf{$\eta(b\bar{b})$:} Pseudorapidity of the two $b$-jets above which exhibits a deeper dip around 0 for background events,   where $b$-jets are scattered back-to-back (signal events with a larger boost show a somewhat different behavior at $\eta(b\bar{b})=0$).
\item \textbf{$\Delta R(b,\bar{b})$:} Angular distance separation between the two $b$-jets reconstructing the Higgs boson, with $\Delta R(b\bar{b}) = \sqrt{(\Delta\eta_{(b\bar{b})})^2+(\Delta\phi_{(b\bar{b})})^2}$. For background events via $ZZ$ or $hZ$ production, the two $b$-jets are from the $Z$ or $h$ boson decay. When the $Z/h$ boson is produced near its peak in the running of $\sqrt{\hat{s}}$, most of the gauge/Higgs boson evens have small boost factors. Then the pair of $b$-jets from $Z/h$ decay are expected to fly back-to-back. A similar behavior also applies to the $b$-jets emerging from top quark decays, for which $\Delta R(b,\bar{b})$ peaks around 3. For the QCD final state, $Zb\bar  b$, the angular separation between the two $b$-jets has a broader peak, as the $b$-jets come from QCD radiation. The signal events, with a with larger boost of the $b$-jets, allow for $\Delta R(b,\bar{b})$ to peak around 1.
\item \textbf{$M(b,\bar{b})$:}  Invariant mass of the two $b$-jets which is    closest to the SM-like Higgs boson mass. For  signal events,  $M(b, \bar{b})$ peaks at the Higgs boson mass while, for the processes $ZZ$ and $hZ$, $M(b,\bar{b})$ peaks at the $Z$ and (somewhat more broadly than the signal) Higgs boson mass, respectively, while, for the $t\bar{t}$ and $Zb\bar  b$ processes, $M(b,\bar{b})$ does not show any obvious peak.

\end{itemize}
To adjust the reconstructed distributions as an input to the DNNs, we stack all background and signal events separately such that each data set has dimensions of $d_{\text{distribution}} = (N,9)$  with $N$ being the total number of events. In our analysis, we use equal size data sets for signal and background events,  $N=200000$ each.  As the network does not understand the meaning of signal or background,  we assign a numeric label $Y=1$ to the former and $Y=0$ to (the whole of) the latter. Once the labels are adjusted  during the training of the network, the model tries to minimize the error between its predictions and the assigned labels for each data set of signal or background events. The network then repeats this process until it reaches the desired classification accuracy. Finally, it is important to mention that a DNN convergence to a global minimum is sensitive to the ranges of the input data sets, this is why we normalize all distributions to one.

%%%%%%%%%%%%%%%%%%%%%%%%%%%%%%%%%%%%%%%%%
\subsection{Color flow in signal and background}%%%
\begin{figure}[h!]
\begin{center}
\includegraphics[scale=0.65]{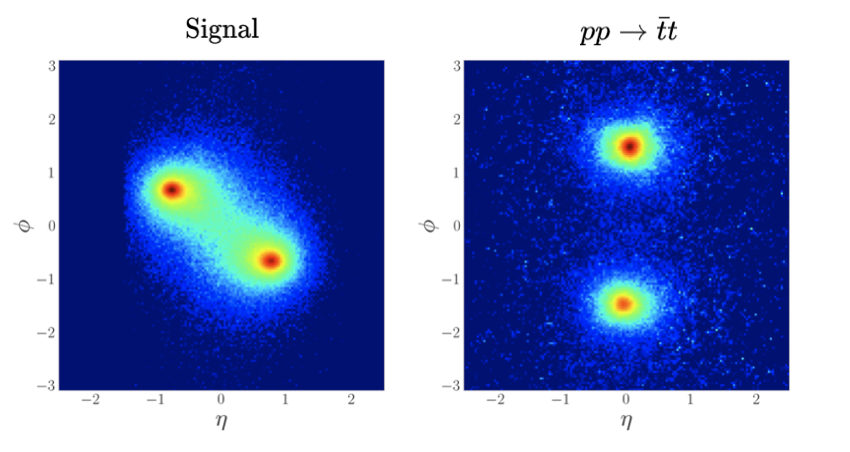}
\caption{Cumulative $p_T$ distributions resulting from showering 10,000 times a single  event at parton level for the signal (left) and $t\bar{t}$ production (right).}
\label{fig:oneshot}
\end{center}
\end{figure}

%%%%%%%%%%%%%%%%%%%%%%%%%%%%%%%%%%%%%%%%%
$SU(3)$ color symmetry implies that color is conserved in the jet interactions providing different color flows for different processes. The structure of the color flow depends on how the group indices are contracted at each vertex, e.g., jets produced from a color singlet have their indices contracted with each other thereby forming a color dipole while jets produced from the decay of a colored particle have their indices contracted with those of the parent particle thereby forming isolated poles connected to the parent particle \cite{Maltoni:2002mq,Hagiwara:2010vk,Kilian:2012pz}. The global structure of the color flow can be seen in LHC events as a color string from the soft hadrons that stretches between two color-connected jets \cite{Gallicchio:2010sw,Kim:2019wns}. This dynamics thus potentially provides with color observables which can be used to aid the search for new physics.
Specific to our case, the two $b$-quarks from the hadronic decay of the Higgs boson form a color dipole whose radiation pattern is contained primarily within a pair of cones around the two $b$-quark directions, with a tendency for more radiation to occur in the region between the two $b$-quarks. In contrast, the two $b$-quarks in the hadronic decay of a colored particle, e.g., a top (anti)quark, are color connected to the incoming proton beams forming two isolated cones with less radiation in the region between the two $b$-quarks. Since color flow is physical, it may be possible to extract the color connections of an event. In Fig.~\ref{fig:oneshot} we show, for illustrative purpose only,  the cumulative transverse momentum  distribution from showering a parton level signal and $t\bar{t}$ event 10,000 times each. As can be seen, indeed, the two $b$-quarks from the Higgs boson decay tend to radiate towards each other allowing for soft radiation to fill the region between them while the $b$-quarks from (anti)top-quark decay are 
color-connected to the incoming beam and thus the soft radiation tends to fill the outer region between the two $b$-quarks. 

\begin{figure}[h!]
\includegraphics[scale=0.2]{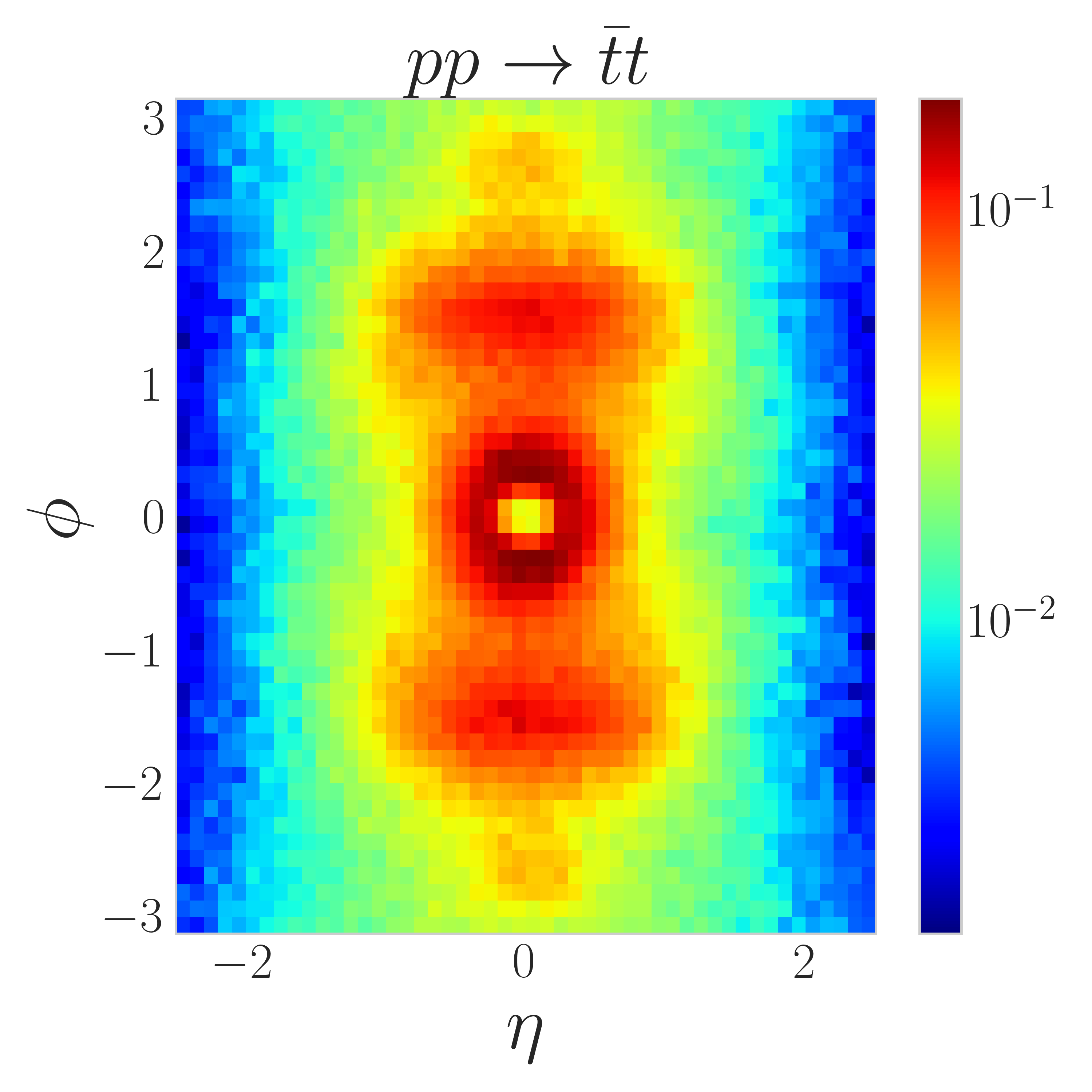}\includegraphics[scale=0.2]{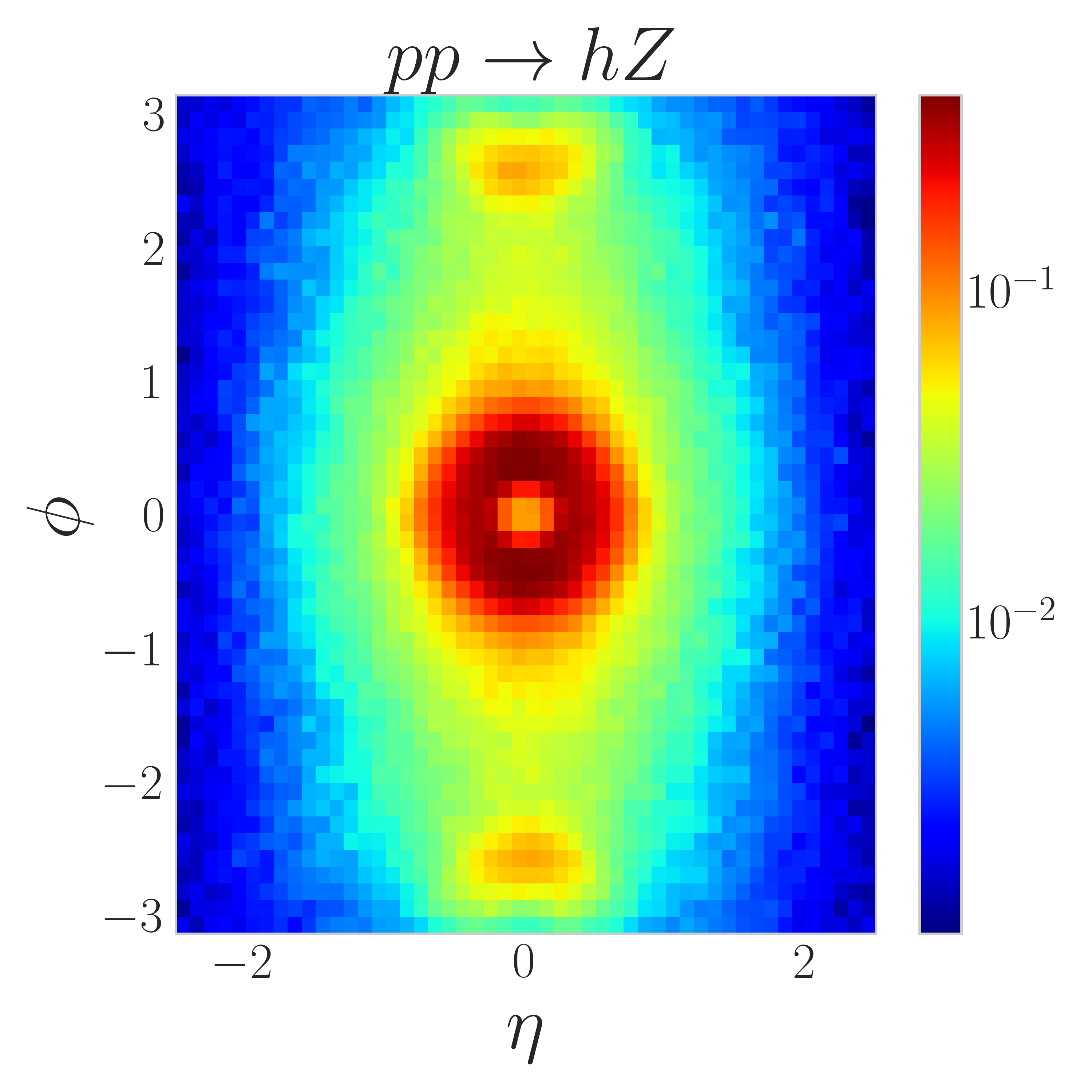}\includegraphics[scale=0.2]{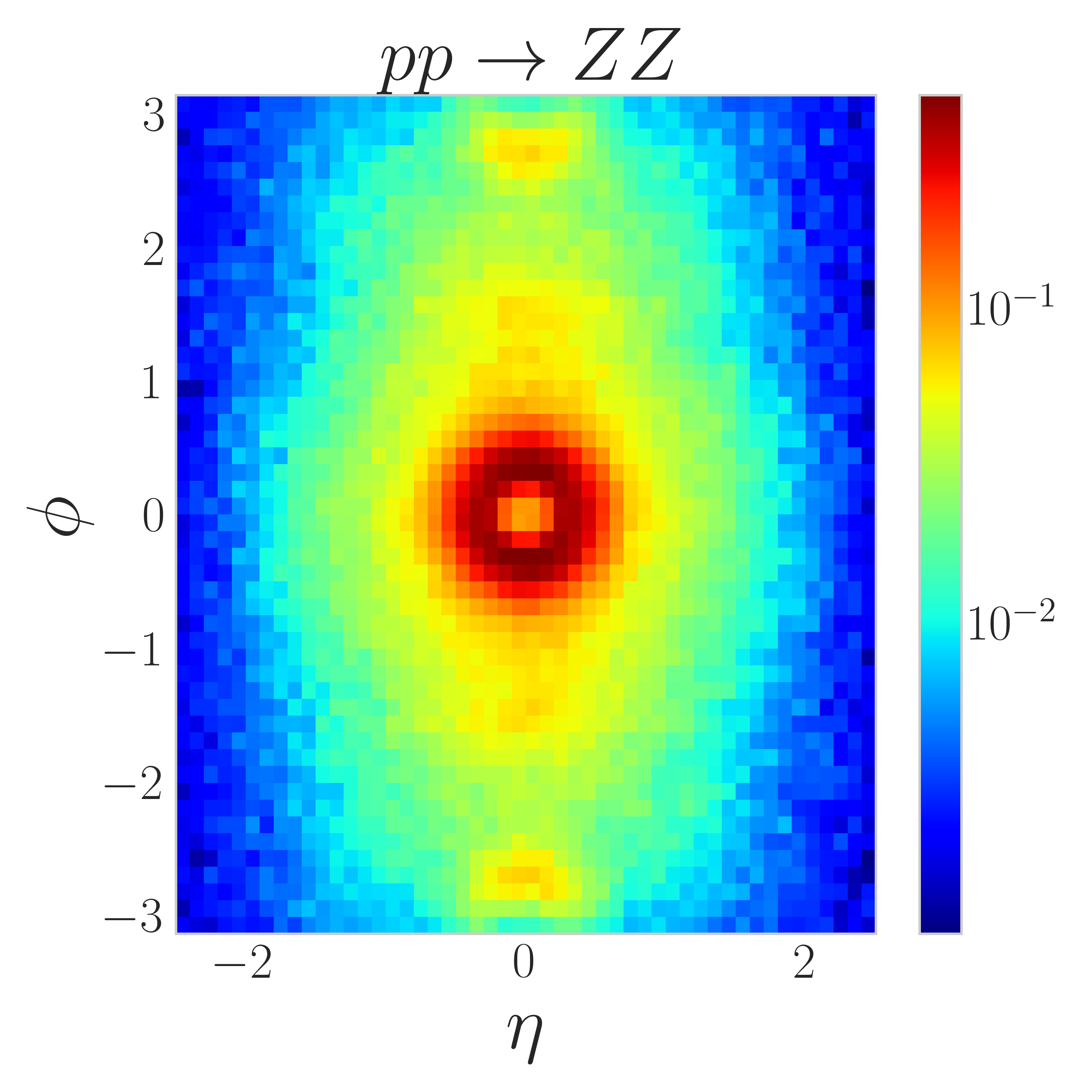}\includegraphics[scale=0.2]{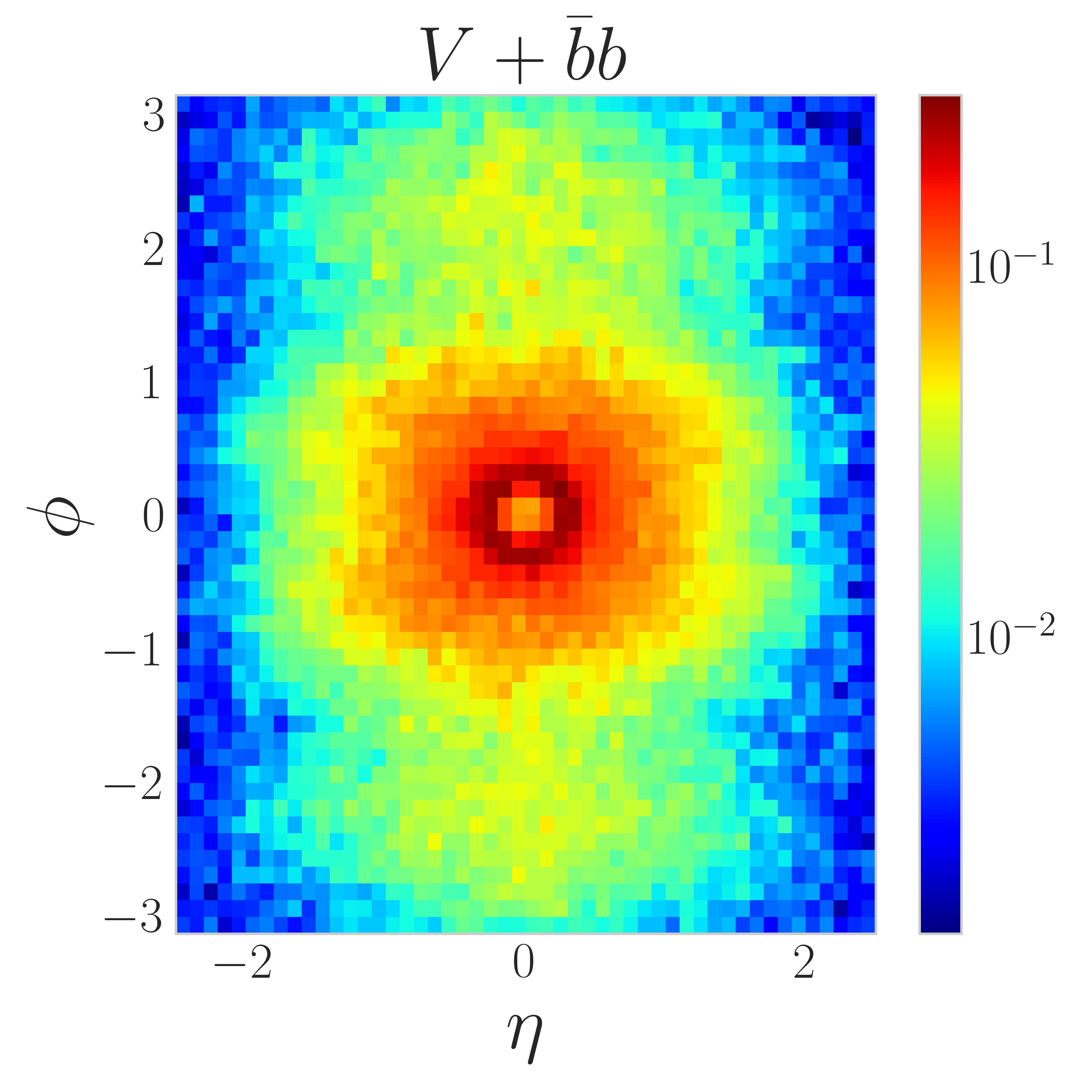}\includegraphics[scale=0.2]{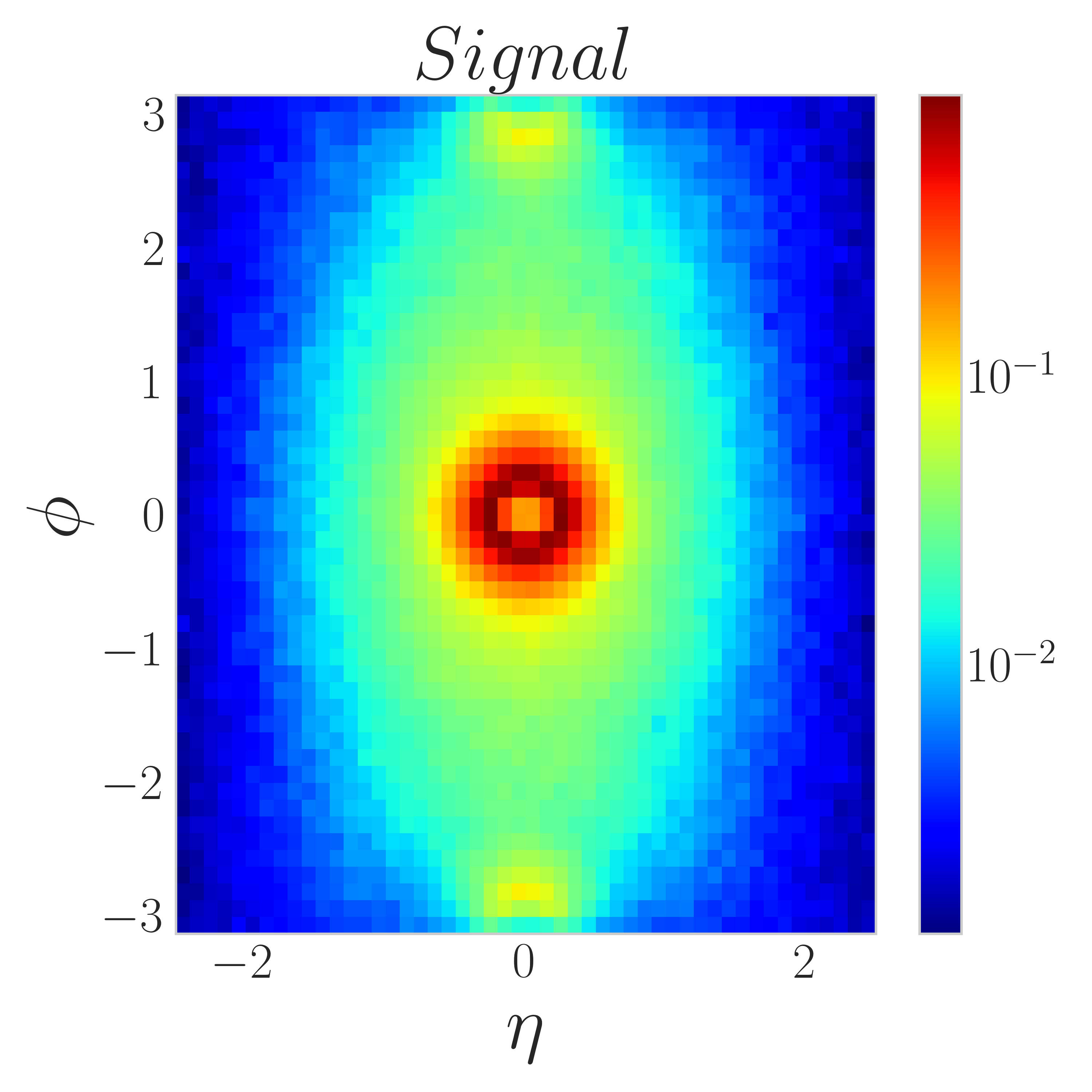}
\caption{Accumulated average of the discussed $p_T$ distribution for 50000 events after a DELPHES simulation for  signal and backgrounds events.}
\label{fig:cumulativeAv}
\end{figure}

 In order to explore the global color structure of actual signal and background events entering our MC analysis, we can use the energy deposit in the LHC calorimeters to depict the information of the final state hadrons into images. Indeed, the calorimeters already provide the requisite pixelization of the constructed image. The pixel intensity of the image can be weighted according to the total transverse momentum of the final state hadrons while the image dimensions will be the pseudorapidity and the azimuthal angle of the calorimetric element.  The results in  Fig.~\ref{fig:oneshot} are only valid in a statistical sense, in fact, since we use  the same parton level event and shower it multiple times. In reality, events are independent and we remark that only one $b$-quark pair is embedded into an image with different locations for it in each event. Thus, in Fig.~\ref{fig:cumulativeAv} we show the accumulated average of the described $p_T$ distribution for 50,000 events (with image pre-processing steps discussed in the next section) after a DELPHES simulation for signal and backgrounds events. 

 To adjust the reconstructed images as an input to the DNNs, we  construct image data sets by embedding each event into an image with dimensions $25\times 25\times 1$. Specifically, we stack all signal and background images separately such that the dimensions of each set are $d_{\text{images}} = (N,25,25,1)$, where $N$ is the number of constructed images for signal and backgrounds, which we take to be $N=20,0000$. In order to improve the network performance and to minimize the ensuing error, it is important to properly process signal and background images before feeding them into the DNNs, as we shall discuss later in the text.

%%%%%%%%%%%%%%%%%%%%%%%%%%%%%%%%%%
\section{Deep learning analysis}\label{sc:4}
%%%%%%%%%%%%%%%%%%%%%%%%%%%%%%%%%%
In this section we discuss  how to construct the aforementioned three independent DNNs  to classify the mono-Higgs signal events from the various backgrounds taken together. We thus discuss the model architecture, pre-processing of input data, model training and classification efficiency. Finally, we compare the classification performance for the three networks.
%%%%%%%%%%%%%%%%%%%%%%%%%%%%%%%%%%%%%
\subsection{Image pre-processing}%%%%
%%%%%%%%%%%%%%%%%%%%%%%%%%%%%%%%%%%%%

For each event we construct an image as a square array in the ($\eta, \phi$) plane with each pixel given by the total hadrons $p_T$ deposited in the associated region in the calorimeter. The rectangular region between $-2.5\le \eta\le 2.5$ and $-\pi\le \phi\le \pi$ is discretized into $25\times 25$ pixels grid. To ensure that the DNN is not learning spacetime symmetries, the jet images are pre-processed as follows.
\begin{enumerate}
\item \textbf{Image cleansing:} We consider only particles which have track information. Also, we remove leptons and photons from the constructed images.
\item \textbf{Pixelization:} We discretize the region in the $(\eta, \phi)$ plane into a $25\times 25$ grid with each pixel weighted by the sum of the transverse momentum in it.
\item \textbf{Centering:} We shift the center of the image from $(0, 0)$ to $( \frac{(\eta_{b} +\eta_{\bar{b}})}{2} , \frac{(\phi_{b}+\phi_{\bar{b}})}{2})$.
\item \textbf{Momentum smearing:} We smear the transverse momentum using a Gaussian function with 3 standard deviation to reduce the number of sparse pixels in the images \cite{Buss:2022lxw}.
\item \textbf{Normalization:}  We normalize the pixel intensity by dividing each pixel in the image by the maximum pixel intensity value.
\end{enumerate}
After these image pre-processing steps, we end up with signal and background image data sets ready to be analyzed  by the networks (herein, with image dimension of $25\times 25\times 1$).
%%%%%%%%%%%%%%%%%%%%%%%%%%%%%%%%%%%%%
\subsection{Network architectures}%%%%%%%%%
\label{sc:4.2} %%%%%%%%%%%%%%%%%%%%%%%%%%
%%%%%%%%%%%%%%%%%%%%%%%%%%%%%%%%%%%%%

As intimated, in this paper, we compare the performance of  three DNNs in distinguishing mono-Higgs signals via $h\to b\bar b$ from various backgrounds. We first use the MLP model that processes the 9 constructed kinematic distributions. The model consists of three fully connected layers with a Rectified Linear Unit (ReLU) activation function and last output layer with two neurons and a soft-max activation function. The number of neurons in the first dense layer is 256, in the second layer is 128 and in the third layer is 64.  To avoid over-training,  we insert a dropout layer after each dense layer with $20\%$ dropout rate. An Adam optimizer, i.e., an algorithm for first order gradient based optimization of stochastic objective functions \cite{Kingma and Ba(2014)}, is used to minimize the loss function with learning rate $\eta=0.001$. 

In analyzing jet images, we use a CNN with four convolution layers, one dense layer and one output layer. The first and second convolution layers have 256 kernels with kernel size 3, a ReLU activation function and stride length of 1 plus, in order to keep the dimensions of the original input images, we use a padding layer.  Third and fourth convolution layers have both 128 kernels with kernel size 3 and a ReLU activation function. After the second and fourth convolution layers we use max-pooling layers with size $2\times 2$ with stride of length 2. After the pooling layer we use a dropout layer with a $30\%$ dropout rate.  The last convolution layer is flattened and projected to one fully connected layer with 64 neurons and a ReLU activation function. The output layer has two neurons and soft-max activation function. Again, the Adam optimizer is used to minimize the loss function with learning rate $\eta=0.001$. 

Finally, a HDNN is constructed by combining the above models without the output layers, rather a   layer is inserted to concatenate the two dense layers from the two models. A fully connected dense  layer with 128 neurons is thus added with a ReLU activation function and a dropout layer with $30\%$ dropout rate is also added. The last output layer consists of two neurons and a soft-max activation function. Finally, the Adam optimizer is used again to minimize the loss function with learning rate $\eta=0.001$. 
%%%%%%%%%%%%%%%%%%%%%%%%%%%%%%%%%%%%%
\subsection{Training the network}%%%%%%%%%%
%%%%%%%%%%%%%%%%%%%%%%%%%%%%%%%%%%%%%
Once the data sets are prepared we train the networks to learn the non-linear relationships between the input data and their labels. For signal events, as mentioned, we assign the label $Y=1$ and for background events we assign 
the label $Y=0$. In order to remove the network dependence on the position  of the signal and background events,  we stack the signal and background events in one data set and shuffle it together with the assigned labels. During the network training stage, during each epoch (defined as number of passes of the entire data sets), the network updates the weights assigned to the neurons for each event via backward propagation of errors.  The network then tries to minimize the error between its predictions and the true labels by reaching a global minimum of some loss function. Finally, the network repeats the process until it reaches the desired accuracy. Once the model is trained, we test it by using completely unseen new data sets to measure the network performance. For MLP and CNN models, we divide the data sets, both kinematic distributions and images capturing color dynamics, into $70\%$ training and $30\%$ testing samples with total size of  $200,000$. For the HDNN model, we use the same training and test data sets, with the same  sizes being now for each stream.  Also, we train the model on equal size data sets for both signal and (whole) background.  The network predictions error is quantified by using the categorical cross entropy loss, given as 
\begin{equation}
\text{Loss} = -\sum_i Y_i \log (\hat{Y}_i)\,,
\end{equation}

with $i=0,1$ for signal and background classes, respectively, and where $Y_i,\hat{Y}_i$ are the true and predicted labels for each class. For all networks, we train the model with 20 epochs with batch size equalling a  $500$ sample. The dimension of the final
output probability, $\hat{Y}$,  is $1\times 2$, $(\mathcal{P}_{sig},\mathcal{P}_{bkg})$, with $\mathcal{P}$ ranging between $[0,1]$. If $\mathcal{P}_{sig} > 0.5\ (\mathcal{P}_{bkg} < 0.5)$, the corresponding event is classified as most likely being a signal event and if $\mathcal{P}_{sig} < 0.5 \ (\mathcal{P}_{bkg} > 0.5)$ the corresponding event is classified as most likely being a background event.
%%%%%%%%%%%%%%%%%%%%%%%%%%%%%%%%%%%%%%%%%%%%%%%%%%%
\subsection{Hyperparameters optimization and cross validation}%
%%%%%%%%%%%%%%%%%%%%%%%%%%%%%%%%%%%%%%%%%%%%%%%%%%%
When creating a ML network one has to define the model architecture which contains a set of parameters which values the network itself  is unable to estimate and thus the user has to fix these. Such parameters are called hyperparameters. The network performance is crucially dependent on the fixed values of these hyperparameters and thus we have to search for those that give the optimal network performance. A traditional way for hyperparmeters optimization is the grid search. Grid search is arguably the most basic hyperparameters tuning method. With this technique, one simply builds a model for each possible combination of all hyperparameter values provided, evaluating each model and selecting the architecture which produces the best results.  Although the grid search method can perfectly fix the hyperparameters values that optimize the network performance, it is very time consuming. Consequently, we use a more economical method called random grid search. Random grid search differs from grid search in that one no longer provides a discrete set of values to explore for each hyperparameter, rather, one provides a statistical distribution for each hyperparameter from which their values can be randomly sampled. To find the hyperparameters values for the networks, like done in \ref{sc:4.2}, we carried out a random grid search over the hyperparameters and fixed their values to the highest network performance. For the MLP, we randomly search for the number of neurons in each layer and dropout rate, with number of neurons from 64 to 256 in each layer and dropout rate from $5\%$ to $35\%$. For the CNN, we search for the kernel numbers in each convolution layer, kernel and pooling size as well as dropout rate,  with number of kernels from 64 to 256 in each layer. For both convolution and pooling kernels size we scan over a list of symmetric kernels with size from $(2,2)$ to $(5,5)$ while we use the same MLP ranges to scan over the dropout rate. For the HDNN, in which we concatenate these two networks thus  leading to larger set of hyperparameters, we just use the optimized  hyperparamters values in each previous network.

The fact that the network performance can depend on the random partitioning of the available data into three sets, e.g.,  training, test and validation set, which can change the network performance if we repeat the training and test steps with new splitting. A solution to this problem is a procedure called cross-validation. In this approach, the data set is split into $k$ smaller sets  and the network is trained on $k-1$ sets while the remaining set is used for validation purposes. The performance measure reported by the cross validation method  is then the average of the values computed in the loop. In our training, we use $k = 5$ for all networks.
%%%%%%%%%%%%%%%%%%%
\section{Results}\label{sc:5}
%%%%%%%%%%%%%%%%%%%
We employ our DNNs to perform the described Mono-Higgs analysis at the HL-LHC with $\sqrt s=14$ TeV in energy and 1000 fb$^{-1}$ of integrated luminosity. The discriminating power of each of the  networks will be a measure of how well the signal and background may be characterised through their different features, all entangled together into several kinematic distributions and color flows. 

\begin{figure}[h!]
\includegraphics[scale=0.3]{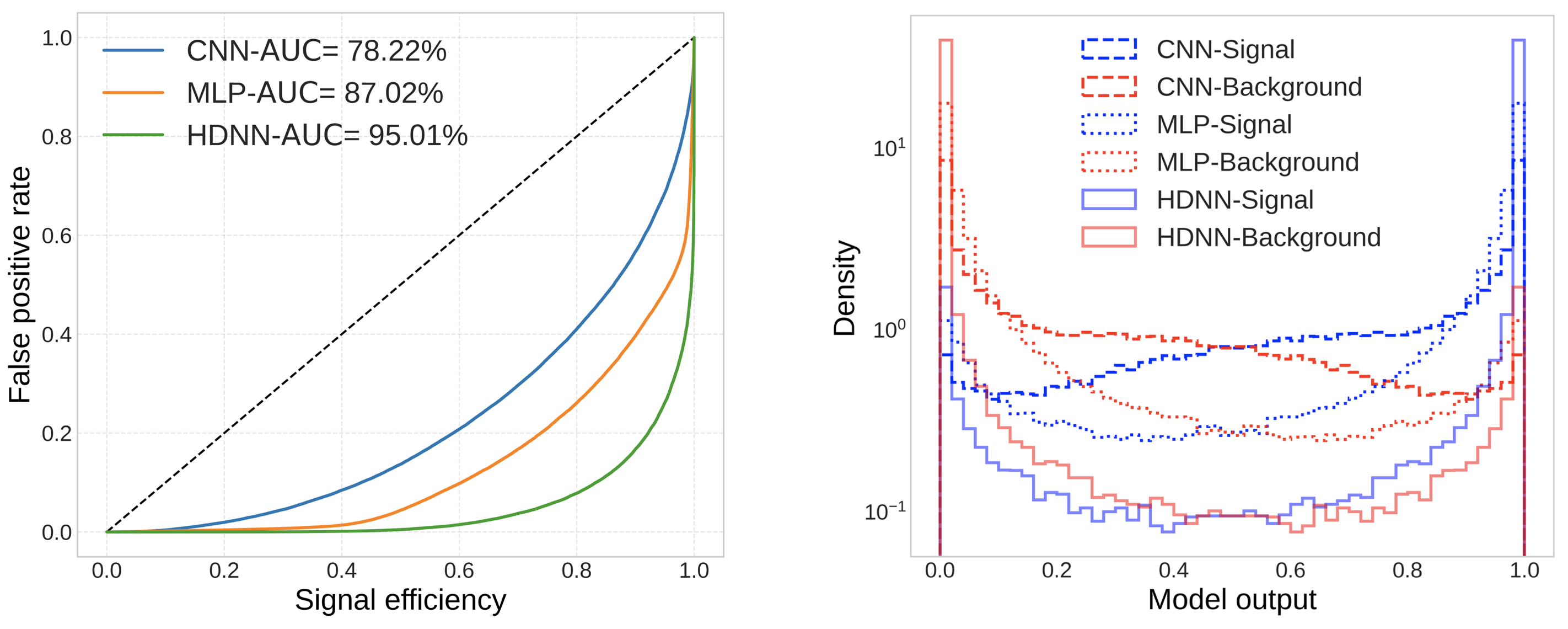}
\caption{Signal efficiency  versus false positive rate with AUC distribution for the three ML types (left).  Model output versus density for signal and background in the three ML types (right).}
\label{fig:ROC}
\end{figure}

Fig.~\ref{fig:ROC} (left) shows the Recursive Operating Curve (ROC) score for the three networks: MLP using kinematic distributions as input, CNN trained on jet images and  HDNN which processes both kinematic distributions and jet images. Herein, the Area Under the Curve (AUC) quantifies the  ability of the network to correctly predict the event class, i.e., signal or background. The MLP gives better performance than the CNN network while the HDNN has the best performance overall for separating the mono-Higgs signal events from the various backgrounds combined. Specifically, the AUC for the HDNN model (green curve) is $96.87\%$ while we find $88.84\% \text{ and } 78.81\%$ for the MLP (orange curve) and CNN (blue curve) network, respectively. 
In Fig.~\ref{fig:ROC} (right) we show the classifier distribution for both signal (blue histograms) and background (red histograms) in the case of the three networks: MLP (short-dashed), CNN (log-dashed) and HDNN (solid), each with the usual inputs. The DNN model output predictions,  $(\mathcal{P}_{sig},1-\mathcal{P}_{bkg})$,  range from 0 to 1: the events with discriminant value near 1 are classified as signal-like events while those near 0 are considered as background-like events. The intersection  area between the two distributions (signal ans background)  indicates the miss-classified events by each network. Again, it is clear that HDNN is the best performer, ahead of MLP and CNN, in turn. 

The optimization of the signal-to-background significance, as a function of the network output, has been performed using the following formula \cite{LHCDarkMatterWorkingGroup:2018ufk,Antusch:2018bgr,Antusch:2020fyz}:
\begin{equation}
\sigma = \left[ 2\left( (N_s+N_b)\ln\frac{(N_s+N_b)(N_b+\sigma^2_b)}{N_b^2+(N_s+N_b)\sigma^2_b}  -\frac{N^2_b}{\sigma^2_b}\ln(1+\frac{\sigma^2_b N_s}{N_b(N_b+\sigma^2_b)})         \right) \right]^{1/2}\,,
\end{equation}
with $N_s$, $N_b$ being the number of signal and background events, respectively, and $\sigma_b$ parameterizing the systematic  uncertainty on the latter. 
\begin{figure}[h!]
\centering
\includegraphics[scale=0.4]{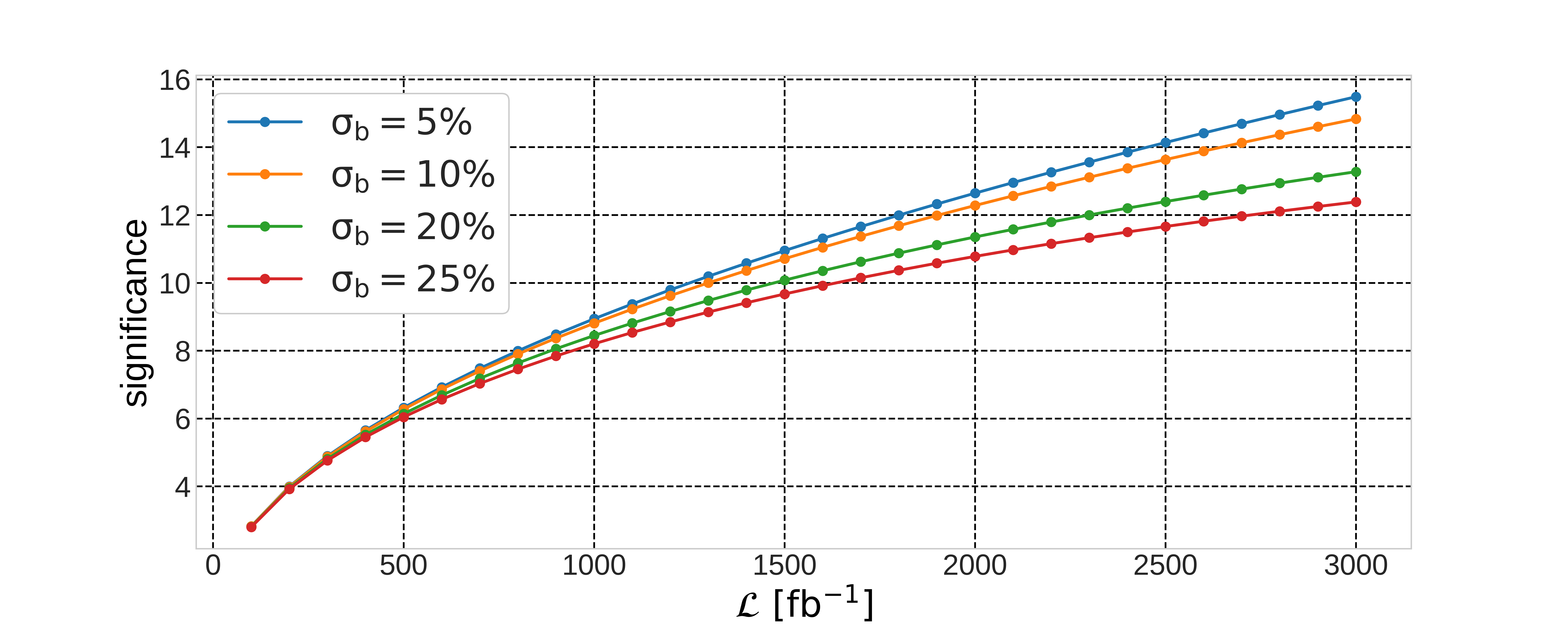}
\caption{Signal significance in terms of integrated luminosity with background uncertainty estimate of $5\%$ (blue), $10\%$ (orange), $20\%$ (green) and $25\%$ (red).}
\label{fig:sign}
\end{figure}
Upon adjusting the cut to maximize the  signal-to-background yield,the CNN shows a best significance of $4.9\sigma$, MLP gives $6.1\sigma$ and HDNN yields $8.9\sigma$. Fig. \ref{fig:sign} shows the HDNN signal significance in terms of the integrated luminosity for different background uncertainties.
%%%%%%%%%%%%%%%%%%%%%%%%%%%%%%%%%%%%
\section{Summary and conclusions}\label{sc:6}

The discovery of a SM-like Higgs boson $h$ with a mass of 125 GeV measured to percent precision by the ATLAS and CMS Collaborations and the extraction of its
$b\bar b$ decays at the LHC has opened up the possibility of using this decay channel as an indirect probe of the existence of DM (and its mediator) in so-called mono-$X$ channels, wherein $X=h$. However, this specific probe suffers from significant QCD background, when standard analysis techniques (i.e., those used in cut-and-count approaches) are used to extract such a signal, so that other $h$ decay channels typically afford one with greater sensitivity in comparison. 

In this paper, we have carried out an analysis which showed the possibility of establishing such a signature in a specific case where the kinematics of the signal is peculiarly different from that of the background, as the DM mediator is a rather heavy object, in the form of a companion Higgs boson existing, e.g., in the BLSSM, which we have used as a template scenario of a Supersymmetry model (wherein the DM candidate has a mass of around 50 GeV). This has been made possible by ML techniques, specifically exploiting HDNNs, which afford one with the
ability of simultaneously using  kinematic distributions and images of jets, the latter obtained by embedding the $p_T$ distribution of the final state particles into visual representations, wherein the different color flow structure of the signal and background processes is accounted for. 

Using such a new computational framework, we have been able to prove the sensitivity of mono-Higgs signals involving $h\to b\bar b$ decays emerging in the BLSSM in presence of large MET over a region of its parameter space comparable to those which have been previously shown to be accessible in such a theoretical framework, specifically involving $h\to\gamma\gamma$ and $ZZ^*\to 4l$ ($l=e,\mu$) decays at the HL-LHC. We regard this as an encouraging result, enabled by ML techniques which one would hope to be also exploitable for other DM production mechanisms where the mediator is lighter, thereby originating less MET in comparison, in turn plagued by a much larger SM background. In fact, eventually, we deem this ML informed approach to be potentially also useful in the search for new Higgs bosons decaying into $b\bar b$ pairs.

%%%%%%%%%%%%%%%%%%%%%%%%%%%%%%%%%%%%
\section*{Acknowledgments}
AH is funded by the grant NRF-2021R1A2C4002551. SM is supported in part through the NExT Institute and the STFC Consolidated
Grant No. ST/L000296/1.

%%%%%%%%%%%%%%%%%%%%%%%%%%%%%%%%%%%%%%

\end{document}